\documentclass{article}
\usepackage{amsfonts}
\usepackage{amsmath}
\usepackage{graphics}
\usepackage{color}

\input{tcilatex}
\usepackage{graphicx}
\begin{document}

\title{Can a charged ring levitate a neutral, polarizable object? Can
Earnshaw's theorem be extended to such objects?}
\author{Stephen J. Minter and Raymond Y. Chiao\thanks{%
Professor in the School of Natural Sciences and in the School of Engineering}
\\
University of California, P. O. Box 2039\\
Merced, CA 95344, U.S.A. \and E-mail: sminter2@ucmerced.edu,
rchiao@ucmerced.edu}
\date{January 22, 2007}
\maketitle

\begin{abstract}
Stable electrostatic levitation and trapping of a neutral, polarizable
object by a charged ring is shown to be theoretically impossible. \
Earnshaw's theorem precludes the existence of such a stable, neutral
particle trap.
\end{abstract}

\section{Introduction\protect\bigskip}

In this tribute in honor of the memory of Prof. Dr. Herbert Walther, we
consider the possibility of extending his famous work on the trapping of an
ordered lattice of ions \cite{Walther-ion-trap} in a Paul trap \cite{Paul},
to the trapping of neutral atoms, and more generally, to the possible
levitation of a macroscopic neutral polarizable object, in a purely
electrostatic trap, for example, in the DC electric field configuration of a
charged ring. \ Earnshaw's theorem will be extended to the case of such
neutral objects, and we shall show below that the stable levitation and
trapping of a neutral, polarizable object, which is a high-field seeker, is
generally impossible in an arbitrary electrostatic field configuration. We
shall do this first for the special case of the electrostatic configuration
of a simple charged ring, and then for the general case of any DC electric
field configuration.\FRAME{ftbpFU}{3.218in}{2.9248in}{0pt}{\Qcb{A uniformly
charged ring with radius $a$ lies on the horizontal $x$-$y$ plane, with its
axis of symmetry pointing along the vertical $z$ axis. Can levitation and
trapping of a neutral particle occur stably near point $L$, where there is a
convergence of $\mathbf{E}$-field lines?}}{\Qlb{Charged-ring-a}}{%
charged-ring-geometry.eps}{\raisebox{-2.9248in}{\includegraphics[height=2.9248in]{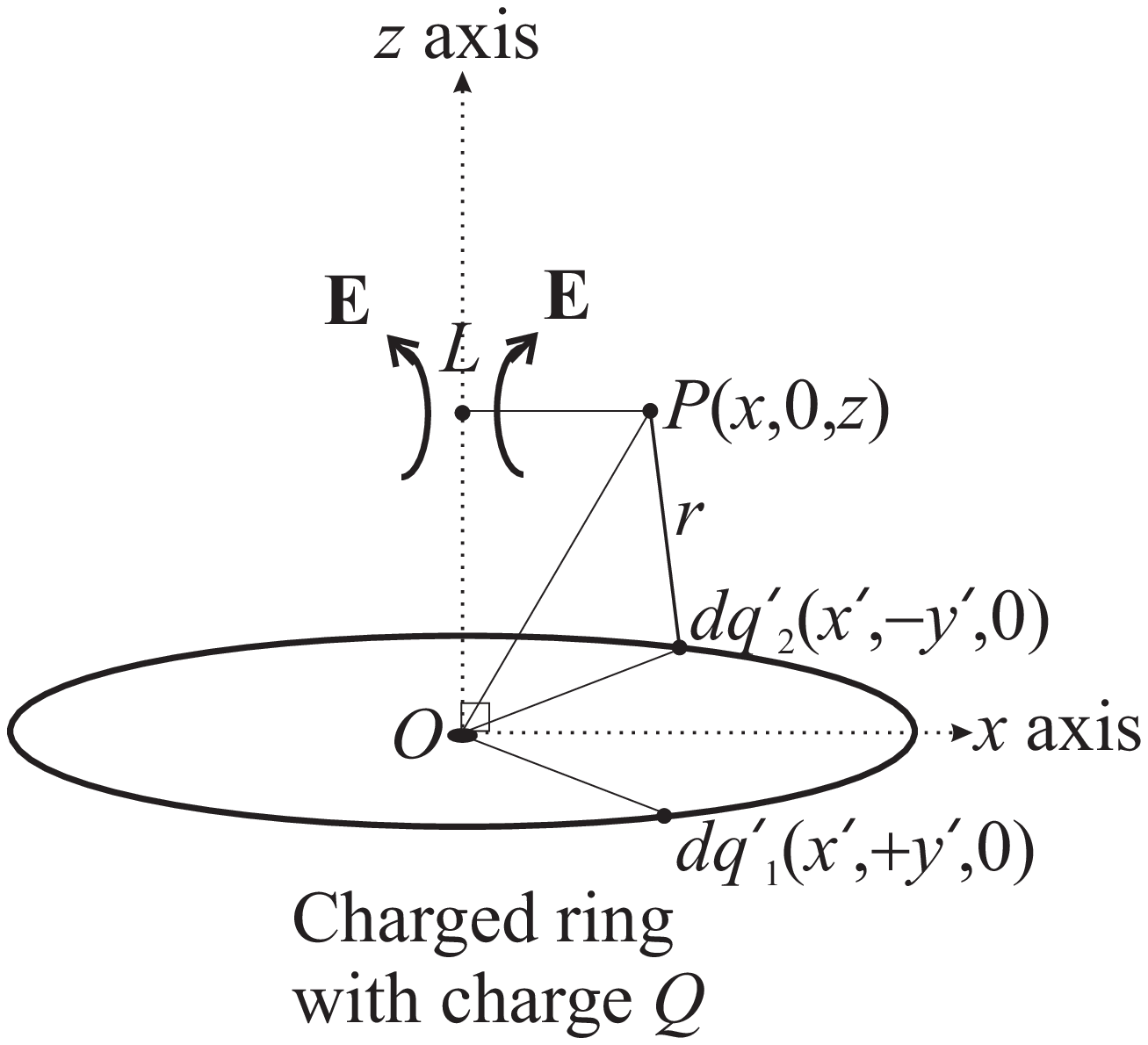}}}

Consider the charged-ring geometry shown in Figure \ref{Charged-ring-a}. The
region near the so-called \textquotedblleft levitation\textquotedblright\
point $L$ in this Figure is akin to the focal region of a lens in optics.
Just as two converging rays of light emerging from a lens in physical optics
cannot truly cross at a focus, but rather will undergo an \textquotedblleft
avoided crossing\textquotedblright\ near the focal point of this lens due to
diffraction,\ so likewise two converging lines of the electric field cannot
cross, and therefore they will also undergo an \textquotedblleft avoided
crossing\textquotedblright\ near $L$.\ There results a maximum in the $z$
component of the electric field along the vertical $z$ axis at point $L$.
The resulting \textquotedblleft avoided crossing\textquotedblright\ region
of electric field lines in the vicinity of point $L$ is therefore similar to
the Gaussian beam-waist region of a focused laser beam. Ashkin and his
colleagues \cite{Ashkin}\ showed that small dielectric particles, which are
high-field seekers, are attracted to, and can be stably trapped at, such
Gaussian beam waists in \textquotedblleft optical tweezers\textquotedblright
. Similarly here a neutral dielectric particle, which is a high-field
seeker, will also be attracted to the region of the convergence of $\mathbf{E%
}$-field lines in the neighborhood of $L$, where there is a local maximum in
the electric field along the $z$ axis.\ The question arises: Can such a
high-field seeker be \emph{stably} levitated and trapped near $L$?

\section{Calculation of the electric potential and field of a charged ring}

The electric potential at the field point $P$ due to a charge element $%
dq^{\prime }$ of the ring is given in by%
\begin{equation}
d\Phi =\frac{dq^{\prime }}{r}  \label{d-Phi}
\end{equation}%
where the distance $r$ from the source point, whose coordinates are $%
(x^{\prime },y^{\prime },0)$, to the field point $P$, whose coordinates are $%
(x,0,z)$, is%
\begin{equation}
r=\sqrt{(x^{\prime }-x)^{2}+y^{\prime 2}+z^{2}}\text{.}  \label{r}
\end{equation}%
(Primed quantities refer to the source point; unprimed ones to the field
point). Since the charged ring forms a circle of radius $a$ which lies on
the horizontal $x$-$y$ plane,%
\begin{equation}
x^{\prime 2}+y^{\prime 2}=a^{2\text{ }}\text{.}  \label{a^2}
\end{equation}

An infinitesimal charge element $dq^{\prime }$\ spanning an infinitesimal
azimuthal angle of $d\phi ^{\prime }$ can be expressed as follows:%
\begin{equation}
dq^{\prime }=\left( \frac{Q}{2\pi a}\right) ad\phi ^{\prime }=-\left( \frac{Q%
}{2\pi a}\right) \frac{ad\left( x^{\prime }/a\right) }{\sqrt{1-(x^{\prime
}/a)^{2}}}  \label{dq}
\end{equation}%
where $Q$ is the total charge of the ring. Let us introduce the
dimensionless variables%
\begin{equation}
\xi ^{\prime }\equiv \frac{x^{\prime }}{a},~\eta ^{\prime }\equiv \frac{%
y^{\prime }}{a},\text{ }\zeta \equiv \frac{z}{a},\text{ }\varepsilon \equiv 
\frac{x}{a}\text{.}
\end{equation}%
Thus%
\begin{equation*}
dq^{\prime }=-\frac{Q}{2\pi }\frac{d\xi ^{\prime }}{\sqrt{1-\xi ^{\prime 2}}}%
\text{ .}
\end{equation*}

Due to the bilateral symmetry of the ring under the reflection $y^{\prime
}\rightarrow -y^{\prime }$, it is useful to sum up in pairs the contribution
to the electric potential from symmetric pairs of charge elements, such as $%
dq_{1}^{\prime }$ and $dq_{2}^{\prime }$ with coordinates $(x^{\prime
},+y^{\prime },0)$ and $(x^{\prime },-y^{\prime },0)$, respectively, shown
in Figure \ref{Charged-ring-a}. These two charge elements contribute equally
to the electric potential $\Phi $ if they span the same infinitesimal
azimuthal angle $d\phi ^{\prime }$. Thus one obtains%
\begin{equation}
\Phi \left( \varepsilon ,\zeta \right) =\frac{Q}{\pi a}\int_{-1}^{+1}d\xi
^{\prime }\frac{1}{\sqrt{1-\xi ^{\prime 2}}}\frac{1}{\sqrt{\varepsilon
^{2}-2\varepsilon \xi ^{\prime }+1+\zeta ^{2}}}\text{ .}
\end{equation}%
Along the $z$\ axis, this reduces to the well-known result%
\begin{equation}
\Phi \left( \varepsilon =0,\zeta \right) =\frac{Q}{a}\frac{1}{\sqrt{1+\zeta
^{2}}}=\frac{Q}{\sqrt{z^{2}+a^{2}}}\text{ .}
\end{equation}%
The $z$ component of the electric field, which is the dominant $E$-field
component in the neighborhood of point $L$, is given by%
\begin{equation}
E_{z}=-\frac{\partial \Phi }{\partial z}=\frac{Q}{\pi a^{2}}\zeta
\int_{-1}^{+1}d\xi ^{\prime }\frac{1}{\sqrt{1-\xi ^{\prime 2}}}\frac{1}{%
\left( \sqrt{\varepsilon ^{2}-2\varepsilon \xi ^{\prime }+1+\zeta ^{2}}%
\right) ^{3}}\text{ .}
\end{equation}%
Along the $z$\ axis, this also reduces to the well-known result%
\begin{equation}
E_{z}=\frac{Qz}{\left( z^{2}+a^{2}\right) ^{3/2}}\text{ ,}
\end{equation}%
which has a maximum value at%
\begin{equation}
z_{0}=\frac{a}{\sqrt{2}}\text{ or }\zeta _{0}=\frac{1}{\sqrt{2}}\text{ .}
\end{equation}%
The \textquotedblleft levitation\textquotedblright\ point $L$ then has the
coordinates%
\begin{equation}
L\left( 0,0,\frac{a}{\sqrt{2}}\right) \text{ ,}
\end{equation}%
neglecting for the moment the downwards displacement of a light particle due
to gravity.

\bigskip The potential energy $U$ for trapping a neutral particle with
polarizability $\alpha $ in the presence of an electric field $%
(E_{x},E_{y},E_{z})$ is given by%
\begin{equation}
U=-\frac{1}{2}\alpha \left( E_{x}^{2}+E_{y}^{2}+E_{z}^{2}\right) \text{ }%
\approx -\frac{1}{2}\alpha E_{z}^{2}\text{,}
\end{equation}%
since the contributions to $U$ from the $x$ and $y$ components of the
electric field, which vanish as $\varepsilon ^{4}$ near the $z$ axis for
small $\varepsilon $, can be neglected in a small neighborhood of $L$.

We now calculate the curvature at the bottom of the potential-energy well $U$
along the longitudinal $z$ axis, and also along the transverse $x$ axis. \
The force on the particle is given by 
\begin{equation}
\mathbf{F}=-\mathbf{\nabla }U\text{ .}
\end{equation}%
Therefore the $z$ component of the force is, to a good approximation,%
\begin{equation}
F_{z}=\alpha E_{z}\frac{\partial E_{z}}{\partial z}\text{ ,}
\end{equation}%
and the Hooke's law constant $k_{z}$\ in the longitudinal $z$\ direction is
given by%
\begin{equation}
k_{z}=-\frac{\partial F_{z}}{\partial z}=-\alpha \left\{ \left( \frac{%
\partial E_{z}}{\partial z}\right) ^{2}+E_{z}\frac{\partial ^{2}E_{z}}{%
\partial z^{2}}\right\} ~,
\end{equation}%
where all quantities are to be evaluated at $L$ where $\varepsilon =0$ and $%
\zeta _{0}=1/\sqrt{2}$. Taking the indicated derivatives and evaluating them
at $L$, one obtains%
\begin{equation}
\left. k_{z}\right\vert _{L}=+\frac{32}{81}\frac{\alpha Q^{2}}{a^{6}}\text{ ,%
}  \label{k_z}
\end{equation}%
where the positive sign indicates a longitudinal stability of the trap in
the vertical $z$ direction.

The $x$ component of the force is, to the same approximation,%
\begin{equation}
F_{x}=\alpha E_{z}\frac{\partial E_{z}}{\partial x}\text{ , }
\end{equation}%
and the Hooke's law constant $k_{x}$\ in the transverse $x$\ direction is%
\begin{equation}
k_{x}=-\frac{\partial F_{x}}{\partial x}=-\alpha \left\{ \left( \frac{%
\partial E_{z}}{\partial x}\right) ^{2}+E_{z}\frac{\partial ^{2}E_{z}}{%
\partial x^{2}}\right\} \text{ ,}
\end{equation}%
where again all quantities are to be evaluated at $L$ where $\varepsilon =0$
and $\zeta _{0}=1/\sqrt{2}$. Again taking the indicated derivatives and
evaluating them at $L$, one obtains

\begin{equation}
\left. k_{x}\right\vert _{L}=-\frac{16}{81}\frac{\alpha Q^{2}}{a^{6}}\text{ ,%
}  \label{k_x}
\end{equation}%
where the negative sign indicates a transverse instability in the horizontal 
$x$ direction.

Similarly, the Hooke's law constant $k_{y}$\ in the transverse $y$\
direction is%
\begin{equation}
\left. k_{y}\right\vert _{L}=-\frac{16}{81}\frac{\alpha Q^{2}}{a^{6}}\text{ ,%
}  \label{k_y}
\end{equation}%
where the negative sign indicates a transverse instability in the horizontal 
$y$ direction. Note that the trap is azimuthally symmetric around the
vertical axis, so that the $x$ and $y$ directions are equivalent to each
other. Because of the negativity of two of the three Hooke's constants $%
k_{x} $, $k_{y}$, and $k_{z}$, the trap will be unstable for small
displacements in two of the three spatial dimensions near $L$, and hence $L$
is a saddle point. Note also that the sum of the three Hooke's constants in
Equations (\ref{k_z}),(\ref{k_x}), and (\ref{k_y}) is zero, i.e., 
\begin{equation}
k_{x}+k_{y}+k_{z}=0.  \label{sum-of-Hooke's-constants}
\end{equation}

\section{\protect\bigskip Earnshaw's theorem revisited}

We shall see that Equation (\ref{sum-of-Hooke's-constants}) can be derived
from Earnshaw's theorem when one generalizes this theorem from the case of a
charged particle to the case of a neutral, polarizable particle in an
arbitrary DC electrostatic field configuration. \ A quantitative
consideration of the force on the particle due to the uniform gravitational
field of the Earth, in conjunction with the force due to the DC
electrostatic field configuration, does not change the general conclusion
that the mechanical equilibrium for both charged and neutral polarizable
particles is unstable.

\section{Charged particle case}

\bigskip We shall first briefly review here Earnshaw's theorem \cite%
{Stratton}, which implies an instability of a charged particle placed into
any configuration of electrostatic fields in a charge-free region of space
in the absence of gravity. Suppose that there exist a point $L$ of
mechanical equilibrium of a charged particle with charge $q$ in the presence
of arbitrary DC electrostatic fields in empty space. \ The potential $\Phi $
for these fields obey Laplace's equation%
\begin{equation}
\nabla ^{2}\Phi =\frac{\partial ^{2}\Phi }{\partial x^{2}}+\frac{\partial
^{2}\Phi }{\partial y^{2}}+\frac{\partial ^{2}\Phi }{\partial z^{2}}=0.
\label{Laplace's equation}
\end{equation}%
Now the force on the charged particle is given by%
\begin{equation}
\mathbf{F}=-q\mathbf{\nabla }\Phi =-q\left\{ \mathbf{e}_{x}\frac{\partial
\Phi }{\partial x}+\mathbf{e}_{y}\frac{\partial \Phi }{\partial y}+\mathbf{e}%
_{z}\frac{\partial \Phi }{\partial z}\right\} =(F_{x},F_{y},F_{z})\text{ .}
\end{equation}%
where $\mathbf{e}_{x}$, $\mathbf{e}_{y}$, $\mathbf{e}_{z}$ are the three
unit vectors in the $x$, $y\,$, and $z$ directions, respectively. By
hypothesis, at the point $L$ of mechanical equilibrium%
\begin{equation}
\left. \frac{\partial \Phi }{\partial x}\right\vert _{L}=\left. \frac{%
\partial \Phi }{\partial y}\right\vert _{L}=\left. \frac{\partial \Phi }{%
\partial z}\right\vert _{L}=0\text{ .}
\end{equation}%
\emph{Stable} equilibrium would require all three Hooke's constants $k_{x}$, 
$k_{y}$, and $k_{z}$ at point $L$\ to be positive definite, i.e.,%
\begin{eqnarray}
k_{x} &=&-\left. \frac{\partial F_{x}}{\partial x}\right\vert _{L}=+\left. 
\frac{\partial ^{2}\Phi }{\partial x^{2}}\right\vert _{L}>0 \\
k_{y} &=&-\left. \frac{\partial F_{y}}{\partial x}\right\vert _{L}=+\left. 
\frac{\partial ^{2}\Phi }{\partial y^{2}}\right\vert _{L}>0 \\
k_{z} &=&-\left. \frac{\partial F_{z}}{\partial x}\right\vert _{L}=+\left. 
\frac{\partial ^{2}\Phi }{\partial z^{2}}\right\vert _{L}>0.
\end{eqnarray}%
However, Laplace's equation, Equation (\ref{Laplace's equation}), can be
rewritten as follows:%
\begin{equation}
k_{x}+k_{y}+k_{z}=0,  \label{sum-of-k's}
\end{equation}%
i.e., the sum of the three components of Hooke's constants for the charged
particle must be exactly zero. The simultaneous positivity of all three
Hooke's constants is inconsistent with this, and hence at least one of the
Hooke's constants along one of the three spatial directions must be
negative. Therefore the system is unstable.

\bigskip The azimuthally symmetric field configurations like that of a
charged ring is an important special case. \ Let $z$ be the vertical
symmetry axis of the ring. Suppose that there is stability in the
longitudinal $z$ direction (such as along the $z$ axis above point $L$), so
that%
\begin{equation}
k_{z}>0\text{ .}
\end{equation}%
By symmetry%
\begin{equation}
k_{x}=k_{y}\equiv k_{\bot }
\end{equation}%
so that Equation (\ref{sum-of-k's}) implies that%
\begin{equation}
k_{\bot }=-\frac{1}{2}k_{z}<0\text{ ,}
\end{equation}%
implying instability in the two transverse $x$ and $y$ directions. \ 

Conversely, suppose there is instability in the longitudinal $z$ direction
(such as along the $z$ axis below point $L$), so that%
\begin{equation}
k_{z}<0\text{ .}
\end{equation}%
Again, by symmetry%
\begin{equation}
k_{x}=k_{y}\equiv k_{\bot }
\end{equation}%
so that Equation (\ref{sum-of-k's}) implies that%
\begin{equation}
k_{\bot }=-\frac{1}{2}k_{z}>0\text{ ,}
\end{equation}%
implying stability in the two transverse $x$ and $y$ directions. \ 

\section{Adding a uniform gravitational field such as the Earth's, in the
case of a charged object}

\bigskip The potential energy of a charged, massive particle in a DC
electrostatic field in the presence of Earth's gravitational field is%
\begin{equation}
U_{tot}=q\Phi +mgz.
\end{equation}%
Note that the term due to gravity, i.e., the $mgz$ term, is linear in $z$,
and therefore will vanish upon taking the second partial derivatives of this
term. Therefore the Hooke's constants $k_{x}$, $k_{y}$, and $k_{z}$ will be
unaffected by Earth's gravity. The force on the particle is%
\begin{equation}
\mathbf{F}_{tot}=-\mathbf{\nabla }U_{tot}=-q\mathbf{\nabla }\Phi -mg\mathbf{e%
}_{z}
\end{equation}%
where $\mathbf{e}_{z}$ is the unit vector in the vertical $z$ direction. \
In equilibrium, $\mathbf{F}_{tot}=0$, but this equilibrium is again
unstable, since upon taking another partial derivative of the term $mg%
\mathbf{e}_{z}$ with respect to $z$ will yield zero, and therefore all of
the above Hooke's law constants are the same in the presence as in the
absence of Earth's gravity.

\section{Generalization to the case of a neutral, polarizable particle}

\bigskip Now suppose that there exists a point $L$ of mechanical equilibrium
of the neutral particle with positive polarizability $\alpha >0$\ somewhere
within an arbitrary electrostatic field configuration. Such a particle is a
high-field seeker, and hence point $L$ must be a point of high field
strength. Choose the coordinate system so that the $z$ axis is aligned with
respect to the local dominant electric field at point $L$. Thus the dominant
electric field component at $L$ is thus $E_{z}$. The potential energy $U$
for a neutral particle with polarizability $\alpha $ in the presence of an
electric field $(E_{x},E_{y},E_{z})$ is given by%
\begin{equation}
U=-\frac{1}{2}\alpha \left( E_{x}^{2}+E_{y}^{2}+E_{z}^{2}\right) \text{ }%
\approx -\frac{1}{2}\alpha E_{z}^{2}\text{,}
\end{equation}%
since the contributions to $U$ from the $x$ and $y$ components of the
electric field, which vanish as $\varepsilon ^{4}$ near the $z$ axis for
small $\varepsilon $, can be neglected in a small neighborhood of $L$. The
force on the particle is%
\begin{equation}
\mathbf{F}=-\mathbf{\nabla }U\text{ .}
\end{equation}%
Therefore the $z$ component of the force is, to a good approximation,%
\begin{equation}
F_{z}=\alpha E_{z}\frac{\partial E_{z}}{\partial z}\text{ ,}
\end{equation}%
and the Hooke's law constant $k_{z}$\ in the $z$\ direction is given by%
\begin{equation}
k_{z}=-\frac{\partial F_{z}}{\partial z}=-\alpha \left. \left\{ \left( \frac{%
\partial E_{z}}{\partial z}\right) ^{2}+E_{z}\frac{\partial ^{2}E_{z}}{%
\partial z^{2}}\right\} \right\vert _{L}=-\alpha \left. E_{z}\frac{\partial
^{2}E_{z}}{\partial z^{2}}\right\vert _{L}~,
\end{equation}%
where the last equality follows from the hypothesis of mechanical
equilibrium at point $L$.

Similarly, the $x$ component of the force is, to the same approximation,%
\begin{equation}
F_{x}=\alpha E_{z}\frac{\partial E_{z}}{\partial x}\text{ ,}
\end{equation}%
and the Hooke's law constant $k_{x}$\ in the $x$\ direction is given by%
\begin{equation}
k_{x}=-\frac{\partial F_{z}}{\partial x}=-\alpha \left. \left\{ \left( \frac{%
\partial E_{z}}{\partial x}\right) ^{2}+E_{z}\frac{\partial ^{2}E_{z}}{%
\partial x^{2}}\right\} \right\vert _{L}=-\alpha \left. E_{z}\frac{\partial
^{2}E_{z}}{\partial x^{2}}\right\vert _{L}~,
\end{equation}%
where the last equality follows from the hypothesis of mechanical
equilibrium at point $L$.

Similarly the $y$ component of the force is, to a good approximation,%
\begin{equation}
F_{y}=\alpha E_{z}\frac{\partial E_{z}}{\partial y}\text{ ,}
\end{equation}%
and the Hooke's law constant $k_{y}$\ in the $y$\ direction is given by%
\begin{equation}
k_{y}=-\frac{\partial F_{z}}{\partial y}=-\alpha \left. \left\{ \left( \frac{%
\partial E_{z}}{\partial y}\right) ^{2}+E_{z}\frac{\partial ^{2}E_{z}}{%
\partial y^{2}}\right\} \right\vert _{L}=-\alpha \left. E_{z}\frac{\partial
^{2}E_{z}}{\partial y^{2}}\right\vert _{L}~,
\end{equation}%
where again the last equality follows from the hypothesis of mechanical
equilibrium at point $L$.

Thus the sum of the Hooke's law constants along the $x$, $y$, and $z$ axes
is given by%
\begin{eqnarray}
k_{x}+k_{y}+k_{z} &=&-\alpha \left. \left\{ E_{z}\left( \frac{\partial
^{2}E_{z}}{\partial x^{2}}+\frac{\partial ^{2}E_{z}}{\partial y^{2}}+\frac{%
\partial ^{2}E_{z}}{\partial z^{2}}\right) \right\} \right\vert _{L}  \notag
\\
&=&-\alpha \left. E_{z}\frac{\partial }{\partial z}\left\{ \frac{\partial
^{2}\Phi }{\partial x^{2}}+\frac{\partial ^{2}\Phi }{\partial y^{2}}+\frac{%
\partial ^{2}\Phi }{\partial z^{2}}\right\} \right\vert _{L}=0.
\end{eqnarray}%
Therefore%
\begin{equation}
\left. \left( k_{x}+k_{y}+k_{z}\right) \right\vert _{L}=0\text{ ,}
\label{sum-of-k's-for-neutral-case}
\end{equation}%
and again, the sum of the three Hooke's law constants must be exactly zero
according to Laplace's equation.

Suppose that the system possesses axial symmetry around the $z$ axis with%
\begin{equation}
k_{z}>0\text{ ,}
\end{equation}%
i.e., with stability along the $z$ axis. Then by symmetry%
\begin{equation}
k_{x}=k_{y}\equiv k_{\bot }
\end{equation}%
so that Equation (\ref{sum-of-k's-for-neutral-case}) implies that%
\begin{equation}
k_{\bot }=-\frac{1}{2}k_{z}<0\text{ ,}
\label{transverse-instability-of-neutral-object}
\end{equation}%
implying instability in both $x$ and $y$ directions. This is exactly what we
found by explicit calculation for the case of a neutral, polarizable object
near point $L$ of the charged ring.

\section{Adding a uniform gravitational field such as the Earth's, in the
case of a neutral, polarizable object}

\bigskip The potential energy of a neutral, polarizable, massive particle in
a DC electrostatic field plus Earth's gravity is%
\begin{equation}
U_{tot}=U+mgz.
\end{equation}%
Again, note that the term due to gravity, i.e., the $mgz$ term, is linear in 
$z$, and therefore will vanish upon taking the second partial derivatives of
this term. Therefore again the Hooke's constants $k_{x}$, $k_{y}$, and $%
k_{z} $ will not be affected by Earth's gravity. The force on the particle is%
\begin{equation}
\mathbf{F}_{tot}=-\mathbf{\nabla }U_{tot}=-q\mathbf{\nabla }U-mg\mathbf{e}%
_{z}
\end{equation}%
where $\mathbf{e}_{z}$ is the unit vector in the vertical $z$ direction. \
In equilibrium, $\mathbf{F}_{tot}=0$, but this equilibrium is again
unstable, since upon taking another partial derivative of the term $mg%
\mathbf{e}_{z}$ with respect to $z$ will yield zero, and therefore again all
of the above Hooke's law constants are the same in the presence as in the
absence of Earth's gravity.

\section{Ways to evade Earnshaw's theorem}

Some known ways to evade Earnshaw's theorem and thereby to construct a truly
stable trap for charged or for neutral particles are (1) to use
non-electrostatic fields such as a DC magnetic field (e.g., the Penning trap 
\cite{Penning}) in conjunction with DC electric fields, or (2) to use
time-varying, AC electric fields, rather than DC fields (e.g., the Paul trap 
\cite{Paul}), or (3) to use active feedback to stabilize the neutral
equilibrium of a charged particle in a uniform electric field, such as was
done for a charged superfluid helium drop \cite{Niemela}, or (4) to use the
low-field seeking property of neutral, diamagnetic objects to levitate them
in strong, inhomogeneous magnetic fields \cite{Weilert}. The latter two
methods may be practical for levitating the superfluid helium
\textquotedblleft Millikan oil drops\textquotedblright\ in the experiment
described in \cite{Chiao2006}.

\textbf{Acknowledgments:} We thank Dima Budker, Richard Packard, Kevin
Mitchell, Jay Sharping, and Mario Badal for helpful discussions.

\end{document}